\documentclass[prl,twocolumn,showpacs]{revtex4}
\usepackage{amssymb,amsmath,psfrag,slashed}

\def \beq  {\begin{equation}}
\def \eeq  {\end{equation}}
\def \ba  {\begin{eqnarray}}
\def \ea  {\end{eqnarray}}
\def \non {\nonumber}

\renewcommand{\thefootnote}{\fnsymbol{footnote}}
\begin{document}

\title{Multidimensional Residues for Feynman Integrals with Generic Power of Propagators}

\author{Jian-Hui Zhang}
\email{zhangjh@mppmu.mpg.de}
\affiliation{Center for High Energy Physics, Peking University, 100871 Beijing, China}


%\vspace*{-12mm}
\begin{abstract}
We propose that the concept of multidimensional residues can be used to directly extracting the coefficients of scalar master integrals (with single propagators only) from one-loop Feynman integrals with generic power of propagators. Unlike the usual integration-by-parts (IBP) technique, where one has to solve iteratively a complicated set of equations to carry out the reduction and determine the coefficients of scalar master integrals, using multidimensional residues provides the possibility of directly extracting the coefficients of the master integrals. As the first application of this idea, we show how to directly extract the scalar box integral coefficients.

\end{abstract}

\maketitle

%\newpage
\renewcommand{\thefootnote}{\arabic{footnote}}
\setcounter{page}{1}\setcounter{footnote}{0}

%\section{Introduction}
\noindent
{\bf{Introduction.}}\, Scattering amplitudes are one of the most widely studied quantities within the framework of perturbative quantum field theory. They are important ingredients for computing various observables that can be used to test the theory in which they are defined. Studying scattering amplitudes can also shed light on potential hidden symmetries of the defining theory that are not clearly visible from the Lagrangian. 

From both phenomenological and theoretical perspectives, one needs to study scattering amplitudes beyond tree-level. In general, such amplitudes are written in terms of tensor Feynman integrals, which can then be reduced down to a set of irreducible integrals. At one-loop level, these irreducible integrals comprise scalar integrals with no more than four propagators. This fact has played an important role in recent developments in computing one-loop scattering amplitudes using the so-called on-shell approach~\cite{hep-ph/9403226,hep-ph/9409265,hep-th/0412103,hep-th/0506068,hep-ph/0609007,arXiv:0704.1835,arXiv:hep-ph/0602178} (for recent reviews on this approach, see e.g.~\cite{arXiv:0704.2798,arXiv:0912.3534}). 

The on-shell approach provides an efficient means of constructing the one-loop amplitude from its singularity structures. One can decompose the one-loop amplitude into: the cut-constructible part, i.e. the scalar four-, three-, two- and one-point scalar integral functions multiplied by rational coefficients; and the rational terms. The scalar integral functions contain all (poly)logarithmic dependence of the amplitudes. Hence by computing and matching the branch cuts of the amplitude and of its decomposition in terms of the scalar integral functions, one can extract information on the rational coefficients of the scalar integral functions. For instance, the coefficient of a scalar box integral can be determined by imposing a quadruple cut that uniquely isolates the contribution of this scalar box integral~\cite{hep-th/0412103}. A quadruple cut condition completely freezes the loop momentum (if it is taken as four-dimensional), therefore matching the cuts determines the coefficient of the scalar box integral in a complete algebraic way. Extracting the coefficients of scalar three-, two- and one-point integrals is more complicated, as in these cases cutting the propagators of the desired integral topology does not uniquely isolate the desired coefficient. To determine the desired integral coefficient, one has to disentangle the contribution of other integrals involving not only the cut propagators but also extra propagators. This can be done following e.g. Refs.~\cite{hep-ph/0609007,arXiv:0704.1835}. Since the scalar integral functions are  well-known, after determining all coefficients of these integral functions, one has the result for the cut-constructible part of the amplitude. The second part, i.e. the rational terms can not be computed from four-dimensional cuts, since they are purely rational functions and do not have any cuts. However, when moving to $D$-dimensions, these rational terms also develop branch cuts and can thus be computed from the $D$-dimensional cuts~\cite{hep-th/0506068,hep-ph/0612277}. Another possibility of computing the rational terms is to use on-shell recursion relations inspired by the BCFW recursions~\cite{BCFW} for tree-level amplitudes~\cite{hep-ph/0507005,hep-ph/0604195}. Combining the cut-constructible part and the rational terms, one arrives at a form of the one-loop amplitude in terms of scalar master integrals without performing an explicit tensor reduction.

In Ref.~\cite{hep-ph/0609007} the reduction of one-loop amplitudes was carried out at the integrand level. By investigating the structure of the integrand of a generic one-loop amplitude, the authors of Ref.~\cite{hep-ph/0609007} were able to decompose the integrand of a one-loop amplitude into integrands corresponding to scalar master integrals and the spurious terms, where the spurious terms depend on the loop momentum and yield vanishing contribution upon integration. The coefficients of scalar master integrals as well as of the spurious terms can then be extracted by imposing four, three, two and one on-shell conditions on the integrand of the one-loop amplitude and its decomposition, and then matching them. The rational terms of the one-loop amplitude can also be computed using the method proposed in Ref.~\cite{hep-ph/0609007}.

The above procedure of directly extracting the scalar master integral coefficients applies to generic tensor one-loop Feynman integrals with single propagators. If there is higher power of propagators involved, one has to resort to other techniques. One of the most common such techniques is to use the so-called integration-by-parts (IBP) identities~\cite{171845}. The IBP technique allows one to derive a set of equations relating Feynman integrals with high powers of propagators to those with lower powers. Solving these equations, one is able to reduce Feynman integrals with high powers of propagators. This reduction procedure is an iterative one. By deriving and solving the IBP equations iteratively, one finally achieves a complete reduction of one-loop Feynman integrals with generic power of propagators down to master integrals with single propagators only. The procedure of solving the IBP equations can be rather complicated, and efficient algorithms are usually needed.

In this paper, we propose that the concept of multidimensional residues provides a new efficient means of directly extracting the coefficients of scalar master integrals from one-loop Feynman integrals with generic power of propagators. In contrast to the IBP method, using multidimensional residues allows one to directly determine the coefficients of scalar master integrals without iteratively solving the IBP equations. Multidimensional residues have played an important role in complex analysis and geometry, but are so far not widely used in physics. There have been some physics references discussing their applications in string theory~\cite{hep-th/0304115} and in studying leading singularities of scattering amplitudes in $\mathcal N=4$ super Yang-Mills theory~\cite{multidimresidues}. We argue that when generic power of propagators is involved in a Feynman integral, multidimensional residues have the potential to efficiently extract the coefficients of master integrals. As the first application and test, we show how to directly extract the coefficients of scalar box integrals appearing in the reduction of one-loop Feynman integrals with generic power of propagators, using multidimensional residues.

Before we proceed, we first give a brief introduction to the concept of multidimensional residues.

\vskip .5em

%\section{Multidimensional residues}\label{mdresidues}
\noindent
{\bf{Multidimensional residues.}}\, The discussion in this section is essentially following Ref.~\cite{intromuldimresidues}. 

Suppose one has $n$ holomorphic functions $g_1, g_2, \cdots g_n$ defined in an open set $U\subset\mathbb C^n$, where the $n$ functions have a single common zero $p$ in $U$, one can associate to any holomorphic function $h\in \mathcal O(U)$ the local residue at $p$ of the memomorphic form
\beq
\omega=\frac{h(\mathbf x)}{g_1(\mathbf x) g_2(\mathbf x)\cdots g_n(\mathbf x)} \hspace{.2cm} \mbox{with} \hspace{.2cm} d\mathbf x= dx_1\wedge dx_2\wedge\cdots\wedge dx_n.
\eeq
This then defines a $\mathbb C$-linear operator
\beq
\mathcal O(U)\to \mathbb C, \hspace{.2cm} h\mapsto Res_p\Big(\frac{h(\mathbf x)d\mathbf x}{g_1(\mathbf x)\cdots g_n(\mathbf x)}\Big):=\frac{1}{(2\pi i)^n}\int_{\Gamma_{\mathbf g}(\epsilon)}\hspace{-1em}\omega,
\eeq
where $\Gamma_{\mathbf g}(\epsilon)$ is the real $n$-dimensional cycle
\beq
\Gamma_{\mathbf g}(\epsilon)=\{\mathbf x\in U: |g_i(\mathbf x)|=\epsilon_i, i=1,\dots,n\}
\eeq
with orientation defined by the $n$-form $d\arg(g_1)\wedge\dots\wedge d\arg(g_n)$, and $\epsilon=(\epsilon_1,\dots,\epsilon_n)$ is any $n$-tuple of sufficiently small, positive real numbers.

As a straightforward generalization of the Cauchy formula for one complex variable, the residue of the memomorphic form $\omega$ with $g_i(\mathbf x)=(x_i-p_i)^{a_i+1}$ ($a_i\in \mathbb N$) is given by
\beq\label{highpowerresidue}
Res_p\Big(\frac{h(\mathbf x)d\mathbf x}{\prod_{i=1}^n (x_i-p_i)^{a_i+1}}\Big)=\frac{1}{a_1!\cdots a_n!}\Big(\frac{\partial^{a_1+\cdots+a_n}h}{\partial x_1^{a_1}\cdots\partial x_n^{a_n}}\Big)(p).
\eeq
This relation will be particularly useful when we discuss below the extraction of the coefficients of scalar master integrals arising in the reduction of one-loop Feynman integrals with generic power of propagators.

\vskip .5em

%\section{Scalar one-loop integral coefficients from multidimensional residues}
\noindent
{\bf{Scalar one-loop integral coefficients from multidimensional residues.}}\, To illustrate how the idea of multidimensional residues can be applied to determining the scalar one-loop integral coefficients, let us start from one-loop amplitudes with single propagators only. According to Ref.~\cite{hep-ph/0609007}, the integrand of any $m$-point one-loop amplitude
%\beq
%A(q)=\frac{N(q)}{D_0 D_1\cdots D_{m-1}}, \hspace{1cm}\mbox{with} \hspace{1cm} D_i=(q+p_i)^2-m_i^2
%\eeq
can be decomposed as
\begin{align}\label{spintegranddecomp}
\frac{N(q)}{D_0 D_1\cdots D_{m-1}}&=\sum_{i<j<k<l}^{m-1}\frac{d(ijkl)+\tilde d(q;ijkl)}{D_i D_j D_k D_l}+\non\\
&\hspace{-3em}\sum_{i<j<k}^{m-1}\frac{c(ijk)+\tilde c(q;ijk)}{D_i D_j D_k}+\sum_{i<j}^{m-1}\frac{b(ij)+\tilde b(q;ij)}{D_i D_j}\non\\
&+\sum_i^{m-1}\frac{a(i)+\tilde a(q;i)}{D_i},
\end{align}
where $D_i=(q+p_i)^2-m_i^2$; $d(ijkl), c(ijk), b(ij), a(i)$ are the coefficients of the scalar 4-, 3-, 2-, 1-point integral, respectively; $\tilde d(q;ijkl)$, $\tilde c(q;ijk), \tilde b(q;ij), \tilde a(q;i)$ are the spurious terms. We do not write down the spurious terms without denominators (propagators) in the above expansion, that is, we will always work in a renormalizable theory and choose the renormalizable gauge where such terms vanish. 

As far as the coefficients of scalar one-loop integrals are concerned, we can take the loop momentum as four-dimensional. We choose the same decomposition of the loop momentum as in Ref.~\cite{hep-ph/0609007}
\beq
q^\mu=-p_0^\mu+\sum_{i=1}^4 x_i l_i^\mu,
\eeq
where $l_i^\mu$ are four linearly independent massless momenta constructed from the external momenta. We will view $x_i$ as independent complex variables. The integrand of the amplitude then becomes a function of these complex variables. The scalar box integral coefficient can be determined by computing and matching the residues of both sides of Eq.~(\ref{spintegranddecomp}) at the pole determined by the on-shell conditions
\beq\label{spboxcond}
D_0=D_1=D_2=D_3=0.
\eeq
It is convenient to make a change of variables from $x_i$ to $D_i$ when computing the residue at $D_0=D_1=D_2=D_3=0$. This change of variables introduces a Jacobian
\beq\label{jac}
J_4=\det\Big(\frac{\partial(D_0,D_1,D_2,D_3)}{\partial(x_1,x_2,x_3,x_4)}\Big)
\eeq
into the residue. However, for an amplitude with single propagators, this Jacobian cancels out when matching the residues of both sides of Eq.~(\ref{spintegranddecomp}). The remaining terms yield exactly the same relation between the numerator $N(q)$, the scalar box coefficients and the 4-point like spurious terms as that obtained in Ref.~\cite{hep-ph/0609007}. The relations to determine the scalar triangle, bubble and tadpole coefficients can be obtained by viewing one, two and three of $x_i$ as input parameters, then computing and matching the residues of the amplitude and its decomposition at the poles determined by the on-shell conditions isolating the desired scalar integral coefficients. 

We now argue that the idea of multidimensional residues can also be used to directly determining the coefficients of scalar one-loop integrals arising in the reduction of one-loop Feynman integrals with generic power of propagators. The main point of this argument is that at the poles of the propagators, the integrand of the original integrals with high power of propagators has the same (multidimensional) residues as the integrand of their decomposed form in terms of scalar master integrals with single propagators. As the first application and test of the idea, we show how to directly determine the scalar box integral coefficients using multidimensional residues. As in the reduction of Feynman integrals with single propagators, there exist contributions of spurious terms at the integrand level. The 4-point like spurious terms now look like
\beq
\frac{\tilde d(q;ijkl,h_i h_j h_k h_l)}{D_i^{h_i} D_j^{h_j} D_k^{h_k} D_l^{h_l}},
\eeq 
where the numerator has the same structure as $\tilde d(q;ijkl)$ in Eq.~(\ref{spintegranddecomp}), but now we can have more power of propagators.

Following the idea proposed above, the relation to determine the scalar box coefficient from multidimensional residues is as following:
\begin{align}\label{boxresidue}
Res_p\Big(\frac{N(q)d^4 x}{D_0^{g_0} D_1^{g_1}\cdots D_{m-1}^{g_{m-1}}}\Big)&=Res_p\Big(\sum_{i<j<k<l}^{m-1}\Big(\frac{d(ijkl)}{D_i D_j D_k D_l}\non\\
&\hspace{-4em}+\sum_{h_{i,j,k,l}\le g_{i,j,k,l}}\frac{\tilde d(q;ijkl,h_i h_j h_k h_l)}{D_i^{h_i} D_j^{h_j} D_k^{h_k} D_l^{h_l}}\Big)d^4 x\Big),
\end{align}
where the subscript $p$ denotes the pole determined by the on-shell conditions $D_i=D_j=D_k=D_l=0$. In the following we take $i=0, j=1, k=2, l=3$ as an illustrative example, the procedure for other cases are completely analogous.

As mentioned above, the numerator of the 4-point like spurious terms has the same structure as in Ref.~\cite{hep-ph/0609007}, namely
\beq\label{spuriousterm}
\tilde d(q;0123,h_0 h_1 h_2 h_3)=\tilde d(0123, h_0 h_1 h_2 h_3)T(q),
\eeq
where
\beq
T(q)=\mbox{Tr}[(\slashed q+\slashed p_0)\slashed l_1 \slashed l_2 \slashed k_3\gamma_5],
\eeq
$l_1, l_2$ are two of the four massless momenta constructed from the external momenta as
\begin{align}\label{momdef}
k_1&=l_1+\alpha_1 l_2, \hspace{1cm} k_2=l_2+\alpha_2 l_1, \non\\
l_3^\mu&=\left<l_1|\gamma^\mu|l_2\right], \hspace{1cm} l_4^\mu=\left<l_2|\gamma^\mu|l_1\right]
\end{align}
with $k_i=p_i-p_0$. %We assume the amplitude is written for $p_0=0$ at this point.

We first write the the propagators $D_i (i=0\dots3)$ in terms of the complex variables $x_j (j=1\dots4)$ as
\begin{align}\label{xiintermsofDi}
D_0&=\gamma(x_1 x_2-4x_3 x_4)-d_0,\non\\
D_1-D_0&=d_0-d_1+\gamma(x_1 \alpha_1+x_2),\non\\
D_2-D_0&=d_0-d_2+\gamma(x_2 \alpha_2+x_1),\non\\
D_3-D_0&=d_0-d_3+2[x_1(k_3\cdot l_1)+x_2(k_3\cdot l_2)\non\\
&+x_3(k_3\cdot l_3)+x_4(k_3\cdot l_4)],
\end{align}
where $d_i=m_i^2-k_i^2$. From these equations one can solve $x_j$ in terms of $D_i$ and compute the Jacobian in Eq.~(\ref{jac}). Using Eq.~(\ref{highpowerresidue}), one obtains the residue on the l.h.s. of Eq.~(\ref{boxresidue}) at the pole as
\begin{align}
R_{4p}\big(q_0^\pm\big)&=\frac{1}{\bar g_0!\cdots\bar g_3!}\non\\
&\left.\frac{\partial^{\bar g_0+\cdots+\bar g_3}}{\partial D_0^{\bar g_0}\cdots\partial D_3^{\bar g_3}}\Big(\frac{R_4\big(q_0^\pm(x_i(D_j))\big)}{J_4\big(q_0^\pm(x_i(D_j))\big)}\Big)\right|_{D_{0,1,2,3}=0},
\end{align}
where $\bar g_i=g_i-1$, $q_0^\pm$ correspond to the two solutions to Eq.~(\ref{xiintermsofDi}). $R_4(q_0^\pm)$ is
\beq
R_4(q_0^\pm)=\frac{N(q_0^\pm)}{\prod_{i\ne0,1,2,3}D_i^{g_i}(q_0^\pm)}
\eeq
and $J_4$ is the Jacobian introduced in Eq.~(\ref{jac}).
%\beq
%J_4=\det\Big(\frac{\partial(D_0,D_1,D_2,D_3)}{\partial(x_1,x_2,x_3,x_4)}\Big).
%\eeq
Before computing the residue on the r.h.s. of Eq.~(\ref{boxresidue}), we point out that, by an explicit computation of the Jacobian $J_4$, we find that it has the same $x_i$ dependence as the 4-point like spurious terms. Therefore the ratio $\tilde d(q;0123,h_0 h_1 h_2 h_3)/J_4$ becomes independent of $x_i$ and thus independent of $D_i$. Consequently all spurious terms with any of the propagator power higher than 1 yield vanishing contribution to the residue, since they would involve a derivative with respect to at least one of $D_i$. The only remaining contributions to the residue are
\beq
\frac{d(0123)}{J_4(q_0^\pm)}-\frac{\beta\tilde d(0123,1111)}{4\gamma^2},
\eeq
where $\beta=1/(1-\alpha_1\alpha_2), \gamma=2l_1\cdot l_2$. Note that as the 4-point like spurious term in the case of single propagators, $J_4(q_0^+)=-J_4(q_0^-)$. Matching the residues on both sides of Eq.~(\ref{boxresidue}) then gives the scalar box integral coefficient
\beq\label{boxcoeff}
d(0123)=\left.\frac{J_4(q_0^+)R_{4p}(q_0^+)+J_4(q_0^-)R_{4p}(q_0^-)}{2}\right|_{D_{0,1,2,3}=0},
\eeq
which obviously reproduces the scalar box integral coefficient for the case where all propagators are single propagators. We have checked with several Feynman integral examples involving different power of propagators that, the scalar box integral coefficient computed from Eq.~(\ref{boxcoeff}) exactly agrees with that obtained from the IBP technique~\footnote{We used the Mathematica package FIRE~\cite{arXiv:0807.3243} to solve the IBP equations and extract the scalar box integral coefficients.}. It is reasonable to expect that, as in the case of single propagators, the above procedure of using multidimensional residues to directly extract the scalar integral coefficients continues to work for triangle, bubble and tadpoles, provided that one obtains all the coefficients of spurious terms or finds a way to avoid solving them. We will explore this in a forthcoming paper. Moreover, the idea presented here may apply to high loop level. We will also explore this point in the near future.

%\newpage

%\section{Conclusions}\label{conclusion}
\vskip .5em

\noindent
{\bf{Conclusions.}}\, In this paper we proposed a new efficient means of directly extracting the scalar one-loop integral coefficients from one-loop Feynman integrals with generic power of propagators, using multidimensional residues. In contrast to the IBP technique, which is widely used for the reduction of Feynman integral with repeated propagators, using multidimensional residues allows to directly extract the coefficients of the master integrals without solving iteratively a set of IBP equations. As a first application and test of the idea, we showed how to directly extract the scalar box integral coefficients.

%\section{Acknowledgements}

\end{document}